\documentclass[prl,a4paper,twocolumn,showpacs,showkeys,superscriptaddress,amsmath,amssymb]{revtex4}
\usepackage{graphicx}
\usepackage{dcolumn}
\usepackage{bm}
\usepackage[dvips]{color} 
\begin{document}

\title{Self-localization of magnon Bose-Einstein condensates on the ground and excited levels: from harmonic trap to a box}


\author{S.~Autti}
\affiliation{Low Temperature Laboratory, School of Science and Technology, Aalto University, Finland}

\thanks{Supported by the Academy of Finland, the EU -- FP7 program (\# 228464 Microkelvin), and by the CNRS -- Russian Academy of Sciences collaboration (\# N16569).}

\author{Yu.M.~Bunkov}
\affiliation{Institute N\'{e}el, CNRS, Grenoble, France}

\author{V.B.~Eltsov}
\affiliation{Low Temperature Laboratory, School of Science and Technology, Aalto University, Finland}


\author{P.J. Heikkinen}
\affiliation{Low Temperature Laboratory, School of Science and Technology, Aalto University, Finland}

\author{J.J. Hosio}
\affiliation{Low Temperature Laboratory, School of Science and Technology, Aalto University, Finland}

\author{P. Hunger}
\affiliation{Institute N\'{e}el, CNRS, Grenoble, France}

\author{M.~Krusius}
\affiliation{Low Temperature Laboratory, School of Science and Technology, Aalto University, Finland}

\author{G.E.~Volovik}
\affiliation{Low Temperature Laboratory, School of Science and Technology, Aalto University, Finland}
\affiliation{L.D. Landau Institute for Theoretical Physics, Moscow, Russia}

\date{\today}
\begin{abstract}
  Long-lived coherent spin precession of $^3$He-B at low temperatures
  around $0.2 \, T_\mathrm{c}$ is a manifestation of Bose-Einstein condensation of
  spin-wave excitations or magnons in a magnetic trap which is formed by
  the order-parameter texture and can be manipulated experimentally.  When the number of magnons increases, the
  orbital texture reorients under the influence of the spin-orbit
  interaction and the profile of the trap gradually changes from
  harmonic to a square well, with walls almost impenetrable to magnons. This
  is the first experimental example of Bose condensation in a box. By selective rf pumping the
  trap can be populated with a ground-state condensate or one at any of the excited energy levels.
  In the latter case the ground state is simultaneously populated by relaxation from the
  exited level, forming a system of two coexisting condensates.
\end{abstract}



\maketitle

During the last few years increasing efforts have been invested in the investigation of
Bose-Einstein condensation (BEC) of non-conserved bosons, such as magnons
\cite{BunkovVolovik2007,Dzyapko2011}, photons \cite{Klaers2011}, and
exciton-polaritons \cite{Kammann2011}.  BEC of quasiparticles and
other particle-like excitations is a special case since in thermal
equilibrium their chemical potential vanishes.  Formally BEC
requires conservation of the particle number, but condensation can still be
extended to systems with weakly violated conservation if a dynamic
steady state is created. The loss of (quasi)particles owing
to their decay can be compensated by pumping and thus, for sufficiently
long-lived excitations, the non-zero chemical potential is well defined and
condensation becomes possible. Magnons in superfluid $^3$He-B satisfy this
condition and condensation there is observed as spontaneous long-lived
coherent precession of spins which is accompanied by various phenomena of spin
superfluidity \cite{BunkovVolovik2010}.  Different types of magnon BEC have been identified in $^3$He-B, starting from the so-called ``Homogeneously Precessing
Domain'' (HPD), which represents a bulk condensate state.

In $^3$He-B at low temperatures the HPD state becomes unstable owing to parametric creation of
spin waves, but another type of coherent long-lived NMR signal with several
orders of magnitude smaller amplitude was discovered in this regime
\cite{PS}. The mode was ascribed to magnon condensation in a magnetic trap
formed in a weakly inhomogeneous order parameter distribution or texture \cite{BunkovVolovik2007}, but the piecemeal information obtained with pulsed and cw NMR measurement has been confusing. Here we report the first
measurements with full experimental control of the order parameter texture which
produces the 3D trap. We find that when the
number of magnons increases the profile of the trap changes from harmonic
to a box. The pressure of the
multi-magnon wave function opens a ``cavity'' in a way similar to the
electron bubble in liquid helium. In
quantum field theory such self localization of a bosonic field is known as a
$Q$-ball \cite{Coleman1985}. This texture-free ``cavity'' can be filled by a magnon condensate on any energy level of the trap. For cold atoms in an optical trap, the
formation of a non-ground-state condensate has been discussed
\cite{Bagnato}, but not yet realized. Of great practical importance is the
fact that these magnon condensates can be used to probe the quantum vacuum state of
$^3$He-B in the limit $T \rightarrow 0$ \cite{Eltsov}, where most conventional measuring
signals become insensitive.

\textbf{Magnon condensation:}--There are two approaches to the
thermodynamics of atomic systems: one can fix the particle number ${\cal N}$ or
the chemical potential $\mu$. For magnon condensation, this corresponds to
different experimental situations: to pulsed or continuous wave (cw) NMR,
respectively. In free precession after the tipping pulse, the number of
magnons pumped into the trap is conserved (if losses are neglected). This
corresponds to fixed ${\cal N}$, when the system itself chooses the global
frequency of coherent precession (= the magnon chemical potential $\mu$
\cite{BunkovVolovik2010}). The opposite case is cw NMR, when a small rf
field is continuously applied to compensate for the losses. The frequency
of precession $\omega$ is then that of the rf field $\omega_{\rm rf}$, the
chemical potential is $\mu \equiv \omega = \omega_{\rm rf}$, and the number of magnons adjusts itself to this
frequency, to match the resonance condition.

\begin{figure}[t]
\centerline{\includegraphics[width=0.8\linewidth]{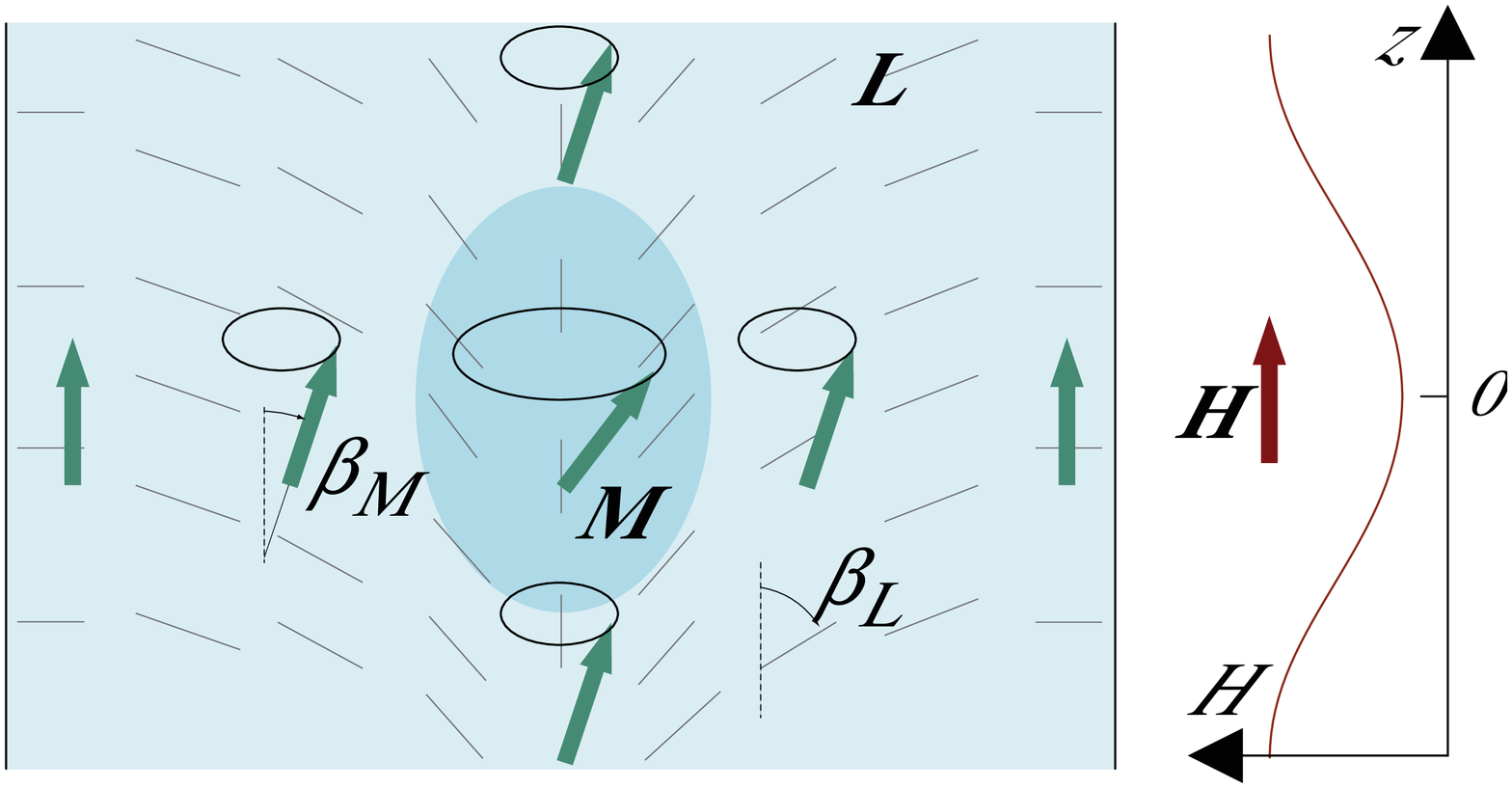}}
 \caption{A sketch of the trapping potential Eq.~\eqref{Potential1} which is formed in the cylindrically symmetric ``flare-out'' texture of the  orbital anisotropy axis $\mathbf{L}$ (thin lines) in a shallow minimum of the vertical magnetic field $\mathbf{H}$ ({\it right}). The arrows represent the magnetization $\mathbf{M}$, which precesses coherently with constant phase angle in the condensate droplet (dark blue), in spite of the inhomogeneity in the texture and in the magnetic field.}
 \label{Fig1}
\end{figure}

A cylindrically symmetric trap for magnons is schematically shown in
Fig.~\ref{Fig1}. It is realized in a long cylindrical sample container with radius $R_\mathrm{s} = 3\,$mm in an
axially oriented magnetic field (the experimental setup is
described in Ref.~\cite{Experiment}). The axial confinement potential  $U_\parallel(z)=\omega_\mathrm{L}(z)$, where $\omega_\mathrm{L}(z)=\gamma H(z)$ is the local Larmor frequency, is  produced by a small pinch coil, which creates a shallow minimum in the magnetic field. The radial confinement comes from the spin-orbit interaction  with the texture of the orbital angular momentum ${\bf L}$ of Cooper pairs \cite{BunkovVolovik2010}:
\begin{equation}
  F_{\rm so}=\frac{4\Omega_\mathrm{L}^2}{5\omega_\mathrm{L}}\sin^2\frac{\beta_\mathrm{L}(r)}{2}
\vert\Psi\vert^2 ,
     \label{FD}
  \end{equation}
  where  $\Psi$ is the wave function of the magnon condensate, $\Omega_\mathrm{L}$ is the Leggett frequency  characterizing the strength of the spin-orbit coupling, and $\beta_\mathrm{L}$ the polar angle of ${\bf L}$.
 The trapping potential is then given as
  \begin{equation}
  U({\bf r})= U_\parallel(z) + U_\perp(r)= \omega_\mathrm{L}(z) +
\frac{4\Omega_\mathrm{L}^2}{5\omega_\mathrm{L}}\sin^2\frac{\beta_\mathrm{L}(r)}{2}~.
\label{Potential1}
\end{equation}
 On the side wall of the cylinder ${\bf L}$ is oriented normal to the wall,  while on the cylinder axis  ${\bf L} \parallel \mathbf{H}$. Close to the axis $\beta_\mathrm{L}$ remains small and varies linearly with distance $r$. Here the potential  $U({\bf r})$ reduces to that of a harmonic trap, as used for the
confinement of dilute Bose gases \cite{PitaevskiiStringari2003},
 \begin{equation}
U({\bf r})= U(0) + \frac{m_\mathrm{M}}{2} \left( \omega_z^2 z^2 + \omega_r^2 r^2\right)~,
\label{Potential2}
\end{equation}
where $m_\mathrm{M}$ is the magnon mass. The low-amplitude standing spin waves have the conventional spectrum
\begin{equation}
 \omega_{\mathrm{mn}}=\omega_\mathrm{L}(0)+   \omega_r(\mathrm{m}+1) + \omega_z(\mathrm{n} +1/2)  \,,
\label{SpinWaveSpectrum}
\end{equation}
where $\omega_\mathrm{L}(0) = \omega_\mathrm{L}(z = 0)$ is the Larmor frequency at the bottom of the well, which corresponds to the center of the trap in Fig.~\ref{Fig1}. The axial oscillator frequency $\omega_z$ is adjusted by changing the current in the pinch coil, while the radial frequency $\omega_r$ can be controlled by rotating the sample, since adding vortex-free superfluid flow or rectilinear vortex lines modifies the flare-out texture. Let us consider the condensates which form when we start filling magnons to one of the levels (m,n) in  Eq.~(\ref{SpinWaveSpectrum}).

\begin{figure}[t]
\centerline{\includegraphics[width=\linewidth]{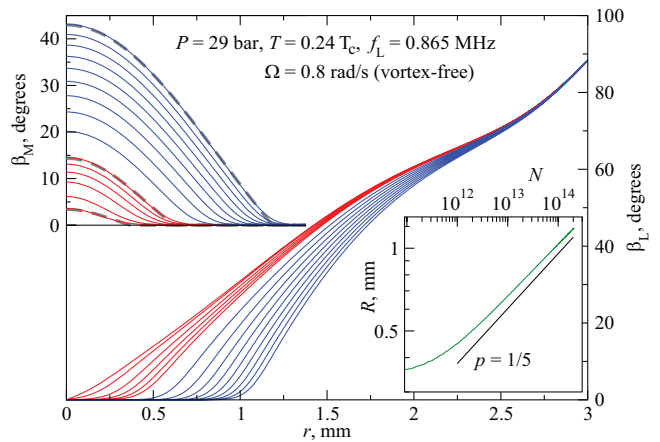}}
\caption{Calculations of multi-magnon bubbles with the condensate in the ground state,
  $n=m=0$. The deflection angles of the magnetization, $\beta_{\rm M}$, and
  of the textural anisotropy axis, $\beta_{\rm L}$, are plotted as a function
  of radius for different condensate
  populations. The population increases from bottom to top in the upper
  plot and from left to right in the lower plot. The red curves correspond to the experimental regime in
  Fig.~\ref{specexamp} (starting with $M_\perp/M_{\rm HPD} = 5\cdot10^{-4}$ and then continuing from $10^{-3}$
  to $5\cdot10^{-3}$ with a step of $10^{-3}$). The blue curves illustrate extrapolations to the
  asymptotic regime $f \rightarrow f_\mathrm{L} (0)$ ($M_\perp/M_{\rm HPD}$
  varies from $10^{-2}$ to $5\cdot10^{-2}$ with a step of $5\cdot10^{-3}$).  When the magnon
  occupation increases, the magnon wave function suppresses the
  orbital texture $\beta_\mathrm{L}$ and the potential well
  transforms towards a box with impenetrable walls. Fits to the wave function of the condensate in a
  box are shown with broken lines in the upper plot. The effective radius
  of the box obtained from such fits is shown in the \textit{inset} as a
  function of the magnon occupation number $\cal N$. The slope from
  Eq.~\eqref{boxrad} with $p=1/5$ is shown for comparison. }
 \label{magnon_bubble}
\end{figure}

\begin{figure}[t]
\centerline{\includegraphics[width=\linewidth]{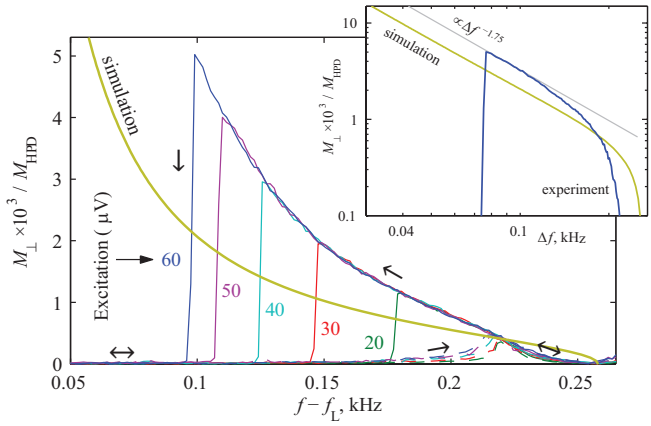}}
\caption{Formation of the magnon condensate in the ground state of
  the trap in Fig.~\ref{Fig1} in cw NMR measurement. The condensate
  magnetization $M_{\perp}$ precessing in the transverse plane is plotted
  on the vertical axis, normalized to that when the \textit{homogeneously
    precessing domain} (HPD) fills the volume within the detector coil \cite{Normalization}. The arrows indicate the sweep direction of the applied rf
  frequency $f = \omega/(2 \pi)$. $M_{\perp}$ grows when the frequency is
  swept down. Only a
  tiny response is obtained on sweeping in the opposite direction. During
  the downward frequency sweep the condensate is destroyed (vertical
  lines), when energy dissipation exceeds the rf pumping. This point
  depends on the applied rf excitation amplitude, marked at each vertical line. The lower green
  line represents the result of calculations from
  Fig.~\ref{magnon_bubble}. These calculations have no fitting parameters:
  the vertical difference from the measurements can be
  attributed to the experimental uncertainty in determining the normalization for $M_{\perp}$. \textit{(Insert)} The experimental curve measured with the
  largest excitation and compared to the numerical curve from the main panel,
  replotted with logarithmic axes to demonstrate the
  asymptotic limit for large magnon numbers:
  $M_{\perp} \propto (f - f_\mathrm{L} (0))^{-1.75}$, which corresponds to
  the condensate in a box.  }
 \label{specexamp}
\end{figure}

\textbf{Ground-state condensate:}--When the number of magnons ${\cal N}$ in
the ground state (0,0) increases, they exert an orienting effect on the ${\bf
  L}$ texture via the spin-orbit interaction in Eq.~\eqref{FD}, which
favors $\mathbf{L}\parallel\mathbf{H}$. As a result at
large ${\cal N}$ the harmonic trap transforms to a box with
$\beta_\mathrm{L}\approx 0$ within which magnons are localized. This effect
is demonstrated in Fig.~\ref{magnon_bubble} with self-consistent
calculations of the texture and of the magnon condensate wave function $\Psi$ in the
axially symmetric and $z$-homogeneous geometry. Rapid flattening of the texture and self localization of the wave function is seen to result. At large ${\cal N}$ the radius of localization approaches the asymptote
\begin{equation}
R({\cal N})\sim ~a_r \left({\cal N}/{\cal N}_\mathrm{c}\right)^{p}~~,~~{\cal N} \gg {\cal N}_\mathrm{c}\,,
\label{boxrad}
\end{equation}
where $a_r$ is the harmonic oscillator length in the original radial trap (at ${\cal N} \ll {\cal N}_\mathrm{c}$), ${\cal N}_\mathrm{c}$ is the characteristic number at which the scaling starts, and $p\approx 0.2$.
As in the case of the electron bubble, $R$ is determined by a balance between the magnon zero-point energy and the surface energy of the condensate bubble. For the 2D radial texture the total energy is:
 \begin{equation}
E(R)={\cal N} \frac{\hbar^2 \lambda_\mathrm{m}^2} {2m_\mathrm{M}R^2} + 2\pi R\sigma(R)\,.
\label{balance}
\end{equation}
The first term on the rhs is the kinetic energy of ${\cal N}$ magnons in a
cylindrical box, where $ \lambda_\mathrm{m}$ is the $m$-th root of the Bessel
function. The second term is the surface energy with the surface tension
$\sigma$ which depends on $R$ due to the flexibility of the texture.  Our numerical
simulations give $\sigma(R)\propto R^2$, and minimization of
Eq.~\eqref{balance} with respect to $R$ gives for the box radius
Eq.~\eqref{boxrad} with $p\approx 1/5$.

\begin{figure}[t]
\centerline{\includegraphics[width=0.43\textwidth]{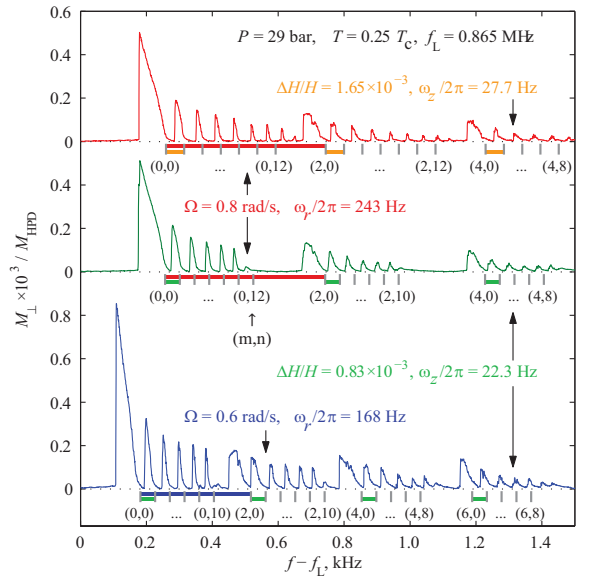}}
 \caption{Formation of magnon condensates at different excited levels (m,n)
   of the magnetic trap in Fig.~\ref{Fig1}. When the frequency of the rf
   excitation field is swept down and crosses one of the levels (m,n) at
   $f_{\mathrm{mn}}$, the  magnon condensate starts to grow according to
   Eq.~\eqref{magnonBEC}.  Only modes with even m and n are excited. The
   three examples of measured spectra are for different values of rotation
   velocity $\Omega$ in the vortex-free state and for different depths
   $\Delta H/H$ of the axial field minimum in the trap. In the topmost
   case the depth of the field minimum is twice that of the two bottom
   examples. As a result, the smaller axial level spacing increases, as the
   axial oscillator frequency $\omega_z/2\pi: \; 22 \rightarrow 27\,$Hz,
   but the larger radial level spacing remains unchanged. Similarly, when
   $\Omega$ is increased from 0.6\,rad/s (bottom spectrum) to 0.8\,rad/s
   (two top most spectra), the radial level separation increases, as
   $\omega_r/2\pi: \; 170 \rightarrow 240\,$Hz. The equidistant vertical
   tick marks refer to the level positions in the absence of magnons and have been fitted
   to the harmonic trap relation in Eq.~(\ref{SpinWaveSpectrum}).
 }
 \label{spectrap}
\end{figure}

Incidentally, for an atomic condensate with repulsive inter-particle interactions the exponent  in Eq.~(\ref{boxrad}) is also $p=1/5$ in the Thomas-Fermi limit \cite{PitaevskiiStringari2003}. However, owing to different scenarios of condensate formation  the  dependence of the frequency shift on ${\cal N}$ differs from the behavior of the analogous quantity in an atomic condensate, \textit{i.e.}, the  chemical potential $\mu( {\cal N} )$ is of the form
\begin{eqnarray}
 \omega -\omega_\mathrm{L}(0)   &\sim&   \omega_r \left( {\cal N}/{{\cal N}_\mathrm{c}}\right)^{-2/5}
~,~~{\rm magnon ~BEC} \, ,
 \label{magnonBEC}
 \\
 \mu - U(0)  &\approx&    \omega_r \left( {\cal N}/{{\cal N}_\mathrm{c}}\right)^{2/5}
 ~,~~{\rm atomic ~BEC}\, .
 \label{atomicBEC}
\end{eqnarray}
In contrast to  condensates of ultra-cold atoms, in the magnon condensate $d\omega/d{\cal N} < 0$. When the magnon condensate is growing, its frequency $\omega$ decreases, approaching the
Larmor frequency $\omega_\mathrm{L}$ asymptotically. This determines the way in which the magnon condensate is grown in a cw NMR experiment (Fig.~\ref{specexamp}).  Magnons are created when the frequency $\omega$ of the applied rf field is swept down and crosses the ground state level $ \omega_{00}$. With increasing  ${\cal N}$,  the potential well becomes wider radially and the energy of the trapped state decreases,
finally  approaching the scaling regime. Assuming that $\sigma(R)\propto R^2$ and taking into account that
${\cal N} =\int d^2r |\Psi|^2$ and thus $ |\Psi| \sim {\cal N}^{1/2} /R$,
one obtains for the transverse magnetization $M_{\perp}\propto \int d^2r
|\Psi| \propto {\cal N}^{1/2} R\propto R^{7/2}\propto (f - f_\mathrm{L}
(0))^{-7/4}$. This agrees with the exponent $ -1.75$ obtained in  numerical
simulations and  in the experiment (Fig.~\ref{specexamp}, insert).

\begin{figure}[t]
\centerline{\includegraphics[width=\linewidth]{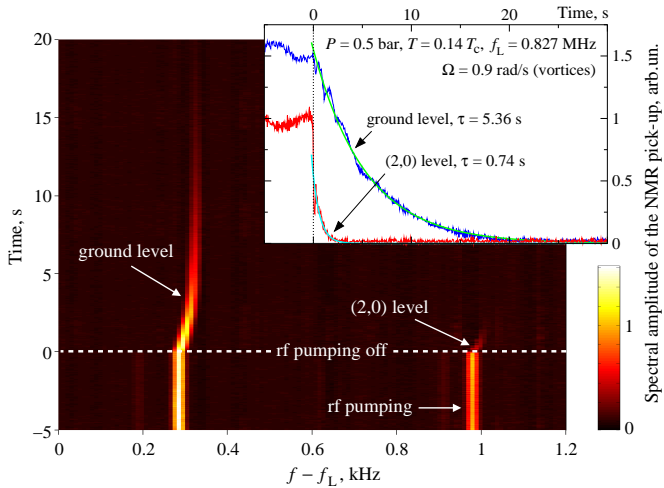}}
\caption{Decay of the magnon condensate, after rf pumping is switched off
  at $t = 0$. The amplitude of the Fourier transform of the signal from the
  NMR pick-up coil is shown in the main panel. Magnons are pumped to the
  (2,0) level at $t<0$, but the ground state is simultaneously populated
  owing to the decay of magnons from the excited state. At $t>0$ both
  states decay and the frequency of precession increases as the trap
  responds to the decreasing magnon population (Fig.~\ref{specexamp}). The \textit{insert} shows the amplitudes of the two peaks and fits to exponential decay with time constant $\tau$. The measurements have been performed at $0.14\,T_\mathrm{c}$ in equilibrium rotation at 0.9\,rad/s with rectilinear vortices.
  }
 \label{decay}
 \vspace{-5mm}
\end{figure}

\textbf{Non-ground-state condensates:}--In similar manner the excited
states $\mathrm{(m,n)}$ can be populated and the entire spectrum
scanned: the condensate in the state $\mathrm{(m,n)}$ starts to grow when
the rf frequency is swept down and crosses the level
$\omega_{\mathrm{mn}}(0)$ from above, since with increasing ${\cal N}$ the
frequency decreases.  In the limit of large magnon occupation ${\cal N}$ in
an excited state, the resulting $Q$-ball is analogous to a bubble in liquid He with an excited
electron.

Three examples of level spectra are shown in Fig.~\ref{spectrap}. The figure
demonstrates how the textural trap is controlled by rotation in the
vortex-free Landau state, where the velocity of the superfluid fraction is $v_\mathrm{s} = 0$. Here the azimuthally flowing superfluid counterflow velocity $v_\mathrm{n} - v_\mathrm{s} =
\vert \mathbf{\Omega} \times \mathbf{r}\vert $
modifies the $\mathbf{L}$-texture, 
making the trap steeper in the radial direction. This increases the
oscillator frequency $\omega_r$ and the spectral distance between the
radial modes (m). Similarly, on increasing the depth of the minimum in the
field $\mathbf{H}$ the distance between the axial modes (n) increases.

The condensates can be maintained in steady state by continuous pumping at
$ \omega_{\mathrm{mn}}({\cal N})$. After switching off the pumping, the
condensate is manifested as long-lived ringing of the free induction
signal.  In Fig.~\ref{decay} the decay of the induction signal has been
recorded, showing two coexisting condensates: at the excited (2,0) level,
where the magnons were initially pumped, and at the ground level (0,0), which is filled
by relaxation from the excited state. This quantum relaxation process is similar to the formation of the magnon condensate with incoherent pumping \cite{Dzyapko2011} and explains
the off-resonance excitation of the ground state population observed in
Ref.~\cite{offreson}.  During the decay the magnon population follows
closely the trajectory for the reverse process of Fig.~\ref{specexamp}.  With
decreasing temperature the relaxation rate in the ground state decreases
rapidly -- life times $\sim 15\,$min have been reported from observations
at the lowest temperatures \cite{Long-LivedRinging}. Our measurements in Fig.~\ref{decay}
on the decay from the excited state (2,0) show faster relaxation.
However, the measured relaxation time is much longer than the dephasing time of the linear
NMR response (about $10\,$ms in the same conditions). This coherence of the
precession is the most striking experimental signature of the Bose-Einstein condensate, not only in the ground
state, but also in the excited states of the trap.

\textbf{Conclusions:}--We have demonstrated the formation of coherently
precessing magnon condensates in a magneto-textural
trap with experimentally controllable potential. As distinct from the traps of ultra-cold atoms, here the trap transforms with increasing magnon number from a harmonic well to a cylindrical box.
Different excitation levels can be selectively populated with condensates. These provide
new opportunities for measurements in the $T \rightarrow 0$ limit \cite{Eltsov}, particularly in terms of their relaxation properties.  An urgent task is to determine
whether the relaxation rate will reveal new information about surface
\cite{Majorana} and vortex-core bound states \cite{CoreStates} which are
Majorana-fermion-like zero-energy modes of the topological insulator superfluid
$^3$He-B.  So far large temperature-independent surface relaxation has been
reported when the magnetic trap borders to a boundary
\cite{SpatialDependence}, while we have preliminary observations of
enhanced relaxation with increasing vortex number within the condensate. The
proper understanding of these effects remains a task for the future.




\begin{thebibliography}{30}
\bibitem{Normalization} Experimentally
$M_\perp = \chi (T) H \int d^3r \, \sin \beta_\mathrm{L}$ is convenient to normalize by means of the HPD where $M_{\mathrm{HPD}} = M_\perp \vert_{\beta_\mathrm{L} = 104^\circ}$, after correcting for the temperature difference.
\end{thebibliography}
\end{document}